\documentclass[conference]{IEEEtran}
\IEEEoverridecommandlockouts
\onecolumn

\AtBeginDocument{%
  \providecommand\BibTeX{{%
    \normalfont B\kern-0.5em{\scshape i\kern-0.25em b}\kern-0.8em\TeX}}}

\usepackage{algorithm}
\usepackage[noend]{algpseudocode} 
\usepackage{algorithmicx}

\usepackage{amsmath,amssymb,amsfonts}
\usepackage{graphicx}
\usepackage[font=small]{caption}
\usepackage{float} 
\usepackage{textcomp}
\usepackage{xcolor}
\usepackage{colortbl}
\usepackage{optidef} 
\usepackage{listings}
\usepackage{lipsum} 
\usepackage[normalem]{ulem}
\usepackage[shortlabels]{enumitem}
\usepackage{subcaption}
\usepackage{graphicx} 
\usepackage{amsmath,amssymb,amsfonts}
\usepackage{multirow}
\usepackage{booktabs}
\usepackage{multicol}
\usepackage{graphicx}
\usepackage{array}
\usepackage{caption}
\usepackage{xcolor,colortbl}
\usepackage{xcolor}
\usepackage{longtable}
\usepackage{ifpdf}
\usepackage{adjustbox}
\usepackage{subcaption}
\usepackage{multirow}  
\usepackage{arydshln}

\definecolor{softgreen}{RGB}{0,128,0}
\definecolor{sdr}{RGB}{254,153,41}

\usepackage{subcaption}
\usepackage[font=small,labelfont=bf]{caption}
\usepackage{graphicx} 
\usepackage[font=small]{caption}
\usepackage{float}   
\definecolor{todo}{RGB}{190, 110, 190}

\begin{document}

\title{Variational Autoencoder-Based Black-Box Adversarial Attack on Collaborative DNN Inference\\
\thanks{This material is based upon work supported by the National Science Foundation (NSF) under Award Numbers: CNS-1943338 and CNS-2401928}}



\author{Shima Yousefi, Motahare Mounesan, Saptarshi Debroy\\ City University of New York\\ Emails: \textit{\{syousefi, mmounesan\}@gradcenter.cuny.edu, saptarshi.debroy@hunter.cuny.edu}}




\maketitle

\begin{abstract}
    In recent years, Deep Neural Networks (DNNs) have become increasingly integral to IoT-based environments, enabling real-time visual computing. 
    However, the limited computational capacity of these devices has motivated the adoption of collaborative DNN inference, where the IoT device offloads part of the inference-related computation to a remote server. 
    Such offloading often requires dynamic DNN partitioning information to be exchanged among the participants over an unsecured network or via relays/hops, leading to novel privacy vulnerabilities.
    In this paper, we propose {\em AdVAR-DNN}, an adversarial variational autoencoder (VAE)-based misclassification attack, leveraging classifiers to detect model information and a VAE to generate untraceable manipulated samples, specifically designed to compromise the collaborative inference process. AdVAR-DNN attack uses the sensitive information exchange vulnerability of collaborative DNN inference and is black-box in nature in terms of having no prior knowledge about the DNN model and how it is partitioned. 
    Our evaluation using the most popular object classification DNNs on the CIFAR-100 dataset demonstrates the effectiveness of AdVAR-DNN in terms of high attack success rate with little to no probability of detection.
    
\end{abstract}




\maketitle

\section{Introduction}

Recent advancements in IoT and edge computing ecosystems have facilitated the deployment of Deep Neural Networks (DNNs) across various applications, including autonomous driving, smart city management, industrial automation, and rescue operations, where real-time decision-making is critical. Consequently, there has been an increased focus on executing DNN inference closer to the data generation site (in contrast to fully remote cloud execution) 
to meet these real-time demands. However, the complexity and resource demand of such DNN models make it infeasible to process them on resource-constrained IoT devices without compromising accuracy. To address this, collaborative inference has been proposed as a solution, distributing the computational workload between IoT devices and remote edge or cloud servers~\cite{zhang2023resource}.


Collaborative inference mandates processing a part of a DNN model (i.e., first few layers) 
on the resource-constrained IoT device. This requires the entire pre-trained DNN model to be preloaded on the IoT device. However, the number of layers to be executed on the device at run-time is dictated by where the DNN is partitioned. Upon execution of these layer at the IoT device, the intermediate output and cut-point information (due to the dynamic nature of where the DNN is partitioned) are transmitted, often over unsecured wireless channels without encryption or through an intermediate node/device, to the remote edge/cloud server to complete the execution of the rest of the layers in order to fully execute the DNN.

Transmitting such privacy-sensitive information over unsecured links and/or potentially malicious intermediate nodes/devices makes the DNNs vulnerable to confidentiality or integrity violations.
Such vulnerability is further compounded due to the difficulty of implementing effective encryption and/or differential privacy measures \cite{zhang2018data} on IoT devices. This is primarily due to the inherent energy and computational constraints of 
IoT devices, and the latency overhead that such an implementation incurs on already  
latency-sensitive DNN inference process. 
In particular, collaborative DNN inference is vulnerable to two main types of confidentiality and integrity violations: data exfiltration, 
in which sensitive information is extracted without authorization; and data falsification, where the data is manipulated or altered. These attacks can target various layers within the edge system, including the application, middleware, and edge layers~\cite{yousefi2024intent}.


One of the primary risks associated with integrity violations of collaborative DNN inference that has raised concerns in recent times is an attacker's ability to manipulate DNNs to induce incorrect classifications \cite{papernot2016limitations}. However, most current research in this space focuses on raw data manipulation, which involves targeting the data even before it is fed into the DNN model. 
While limited work explores data manipulation on the intermediate layers of a DNN during layer partitioning, a soft-spot of such collaborative DNN inference, these studies primarily focus on model reconstruction rather than mitigating misclassifications.
As AI-driven systems become more prevalent, ensuring their security and reliability by addressing vulnerabilities in collaborative inference is essential, as vulnerabilities, such as those in state-of-the-art object detection methods in visual computing, can lead to misclassification attacks, resulting in severe consequences or even critical failures \cite{man2023person}.

In this paper, we investigate the feasibility and impact of data falsification attacks in collaborative DNN inference between IoT devices and edge/cloud servers under varying conditions. 
We first examine the possibility of black-box attacks under these settings and establish their feasibility using detailed benchmarking experiments with state-of-the-art classification DNNs. Building on this, we propose {\em AdVAR-DNN}, an adversarial VAE-based black-box attack capable of manipulating data `in transmit' between the IoT device and the remote server without prior knowledge of the model architecture or precise cut-point specifications. Specifically, AdVAR-DNN attack forms adversarial samples by linear interpolation between the latent representation of the original intermediate data (after the encoder stage of the VAE) and another sample from the training dataset used to train the VAE. This dataset consists of an intermediate representation of the collaborative inference that the attacker can collect through well-known vulnerabilities. The adversarial samples are then fed into the remaining DNN layers at the remote edge/cloud server, leading to incorrect classification.

We evaluated the effectiveness of the proposed AdVAR-DNN attack across different DNN architectures, e.g., AlexNet, VGG19, and MobileNet, in terms of the impact of the attack on the accuracy of DNN models, confidence in misclassifying samples, and the attack success rate under different attack conditions, such as attack scale and attack budget. Our findings demonstrate that different DNN architectures and different cut-point layers within such architectures exhibit distinct vulnerabilities to adversarial perturbations. 
Further, our results show how AdVAR-DNN achieves a high degree of attack success in terms of model confidence in erroneous classification, thus demonstrating a low probability of attack detection. 
Finally, the results show how the chosen linear interpolation technique for latent space manipulation through VAE achieves higher attack effectiveness compared to other alternatives.


The remainder of this paper is organized as follows. Section \ref{sec:Related_Work} discusses the 
related work. 
Section \ref{sec:Feasibility} presents attack feasibility analysis.
Section \ref{sec:attackmodel} discusses the threat model and attack methodology. 
Section \ref{sec:evaluation} describes the evaluation setup and results. 
Section \ref{sec:conclusion} concludes the paper.
\section{Related Work}
\label{sec:Related_Work}
Edge computing enables IoT systems to process a massive amount of data from a wide range of diverse applications closer to the data generation site. 
Such edge adoption introduces new security risks. In \cite{alwarafy2020survey}, the authors explain that an adversary could compromise an edge server or an IoT device that resulting in service manipulation, privacy leakage, and malicious data injection. Additionally, communication between end devices and edge nodes occurs over wired and wireless connections that often lack robust security and make them vulnerable to eavesdropping. \cite{huang2018secure} proposed that an adversary can eavesdrop on wireless communication and analyze traffic patterns to identify a target for further attacks to disrupt normal communication in Edge computing-based Internet of Vehicles. In addition to network risks, machine learning methods used in IoT and edge systems are also vulnerable to attacks during both training 
and inference time. Recent advances have demonstrated that adversaries can leverage adversarial autoencoders to exploit insecure DNN partitioning protocols in collaborative inference settings, enabling reconstruction of both input data and the victim model \cite{manal11073724}.
One can exploit access over the edge infrastructure and recover IoT input image data using power trace collected at the inference stage \cite{wei2018know}. This raises a concern that these vulnerabilities can be used to manipulate model decision-making and bring serious consequences.


One of the most critical security risks in AI/ML IoT applications is adversarial examples resulting in misclassification attacks, which can occur in both physical and digital domains. 
In the physical domain, the adversary could project light onto a camera \cite{man2020ghostimage} or use acoustic signals to interfere with an image stabilization system \cite{ji2021poltergeist}, leading to misclassification.
On the other hand, in the digital domain, perturbations are generated by adding noise to a clean image at the pixel level that is imperceptible to the human eye, while exploiting the generalization learned by DNNs. Su et al. \cite{8601309} proposed an adversarial attack that uses differential evolution (DE) to perturb the value of a single pixel and result in altering the DNN model's prediction, leading to misclassification. \cite{papernot2016limitations} introduced a new algorithm to create adversarial examples by examining how inputs are mapped to outputs in DNNs and applying minor modifications to inputs that cause the model to produce a specific misclassification.

Perturbing input data is considered as an unnatural modification way that may be easily detected. Recently, to mitigate the effects of adversarial perturbations, numerous studies have been conducted that aim to train models with adversarial examples to enhance the model's sensitivity to such threats \cite{cai2021generative}. However, attackers are not always limited to small amounts of adversarial noise applied to the actual data. \cite{song2018constructing} proposed an attack method that generates adversarial samples from scratch using generative models, specifically Auxiliary Classifier Generative Adversarial Networks (AC-GANs). This method can effectively evade defense mechanisms designed to detect noisy adversarial examples \cite{loodaricheh2025handling}. Instead of changing the input data directly, \cite{zhao2017generating} uses a Generative Adversarial Network (GANs) to map latent vectors to real data. By making small changes in the latent space, they generated more meaningful adversarial examples.  

Most adversarial attack studies focus on scenarios where the attacker directly manipulates the input data. However, as mentioned, in real-time applications that depend on collaborative DNN inference, an adversary may have access to the intermediate data exchanged between devices and the central edge server. Some research in this area has focused on the reconstruction of input data from the exchanged information \cite {fredrikson2015model, wei2018know} or model inversion attacks \cite{he2019model} querying the model to reconstruct a shadow model that mimics the behavior of the original model.

\section{Attack Feasibility Analysis}
\label{sec:Feasibility}
Our proposed AdVAR-DNN attack aims to manipulate the intermediate output features of a collaborative DNN inference by mapping them to a latent space, with the ultimate goal of altering the predicted class upon processing the entire DNN. 
In this section, we conduct benchmarking experiments to assess the feasibility of identifying the model and the cut-point layer (by the attacker) using the intermediate output of an unknown layer belonging to an unknown DNN as the sole input.
Additionally, we investigate the potential of a VAE  to manipulate data (using the latent space) in order to trigger misclassification. 
\begin{figure*}[t]
    \centering
    \begin{subfigure}{0.25\linewidth}
        \centering
        \includegraphics[width=\linewidth]{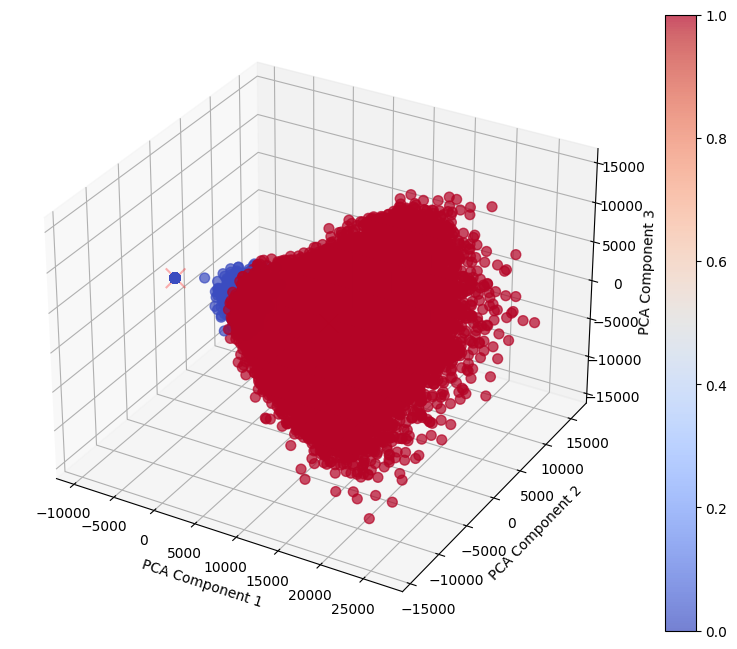}
        \caption{}
        \label{fig:model}
    \end{subfigure}
    \begin{subfigure}{0.25\linewidth}
        \centering
        \includegraphics[width=\linewidth]{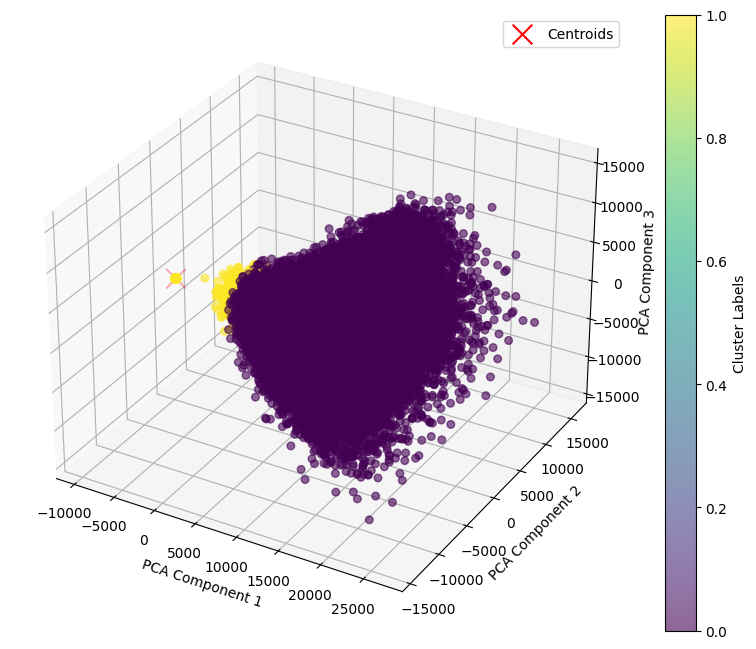}
        \caption{}
        \label{fig:VGG}
    \end{subfigure}
    \begin{subfigure}{0.25\linewidth}
        \centering
        \includegraphics[width=\linewidth]{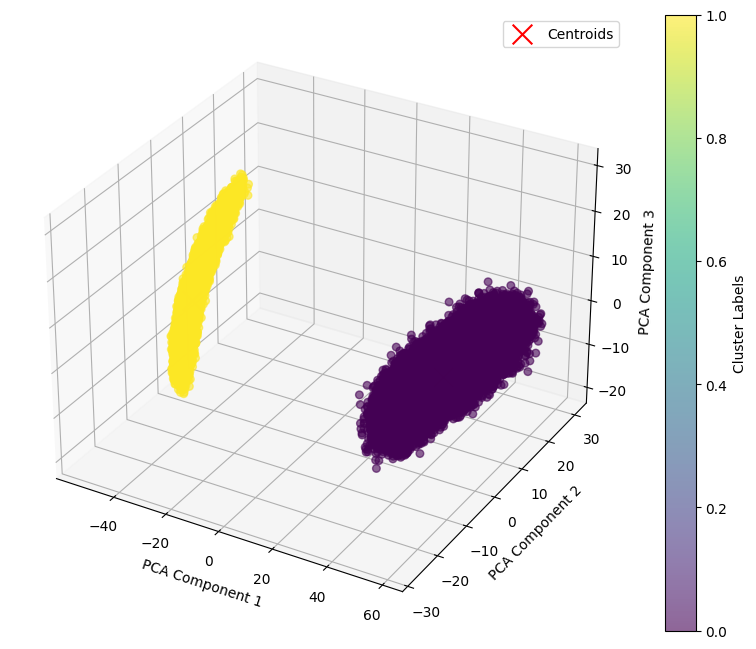}
        \caption{}
        \label{fig:MobiNet}
    \end{subfigure}
    \caption{
    (a) Clustering of features from VGG19 and MobileNet; (b) Clustering of features from two layers of VGG19; (c) Clustering of features from two layers of MobileNet}
    \label{fig:clustering}
    \vspace{-0.2in}
\end{figure*}

\subsection{Model and Cut-point Differentiability}


We speculate that intermediate output data from different models or layers may follow distinct distributions. If these variations produce recognizable patterns in the data, an attacker can distinguish between models, even without knowing the DNN architecture or input data, by clustering them into separate groups.
As a result, they could use this information to manipulate the model accuracy, potentially launching adversarial attacks by generating adversarial samples to replace the original ones.

For these experiments, we examine two well-known object detection DNN models, viz. VGG19 \cite{sengupta2019going} and MobileNet \cite{howard2017mobilenets}. 
Our goal is to determine 
differentiability between the models.
Here, we use the k-means clustering `Elbow method to determine if the intermediate outputs form distinct clusters. This analysis will help us identify potential vulnerabilities in the way model output can be exploited for misclassification attacks. Table \ref{tab:table1} shows the experiment and DNN model details. 

Figure \ref{fig:model} illustrates the k-means clustering results for layers of the same size from the VGG19 and MobileNet models.
The Silhouette Score of 0.881 indicates that the intermediate outputs from these models can be effectively separated into distinct categories. 
The observed imbalance in the scatter plot, where one color represents significantly more points than the other despite a high Silhouette Score, likely indicates that one model's features are more tightly clustered, i.e., dominate the variance in the dataset. This suggests that one model's features are more consistent and form a denser cluster. 

\begin{table}[t]
\centering
\caption{Layer specifications for the benchmarking experiments. 
}
\resizebox{\columnwidth}{!}{
\begin{tabular}{lclc}
    DNN Model & Layer index & Layer name & Layer output size \\
    \hline
    \multirow{2}{*}{VGG19} & 16 & block4\_conv1 & (- , 14, 14, 512) \\
                            & 20 & block5\_conv4 & (- , 14, 14, 512) \\
    \hline
    \multirow{2}{*}{MobileNet} & 40 & conv\_pw\_6 & (- , 14, 14, 512) \\
                               & 63 & conv\_dw\_10\_relu & (- , 14, 14, 512) \\
    \hline
\end{tabular}}
\label{tab:table1}
\vspace{-0.1in}
\end{table}


Next, we aim to observe if the distributions of different DNN layers' outputs can be effectively clustered into distinct groups. To achieve this, 
we focus on the intermediate outputs of two specific layers of VGG19 and MobileNet each, viz., layers 16 and 20 and layers 40 and 63 respectively, which are of the same sizes, as detailed in Table \ref{tab:table1}.
Figure \ref{fig:VGG} illustrates the output of VGG19, which achieves a Silhouette Score of 0.763. This indicates a strong potential for clustering, and the accompanying 3D visualization further demonstrates that the outputs from these layers can be separated into distinct clusters.
Figure \ref{fig:MobiNet} shows the results of 
MobileNet, 
which demonstrate a Silhouette Score of 0.875, indicating a high degree of separation.

\subsection{Differentiability of Output Classes}

\begin{figure}[t]
\centering
\begin{subfigure}{0.49\linewidth}
    \centering
    \includegraphics[width=\linewidth]{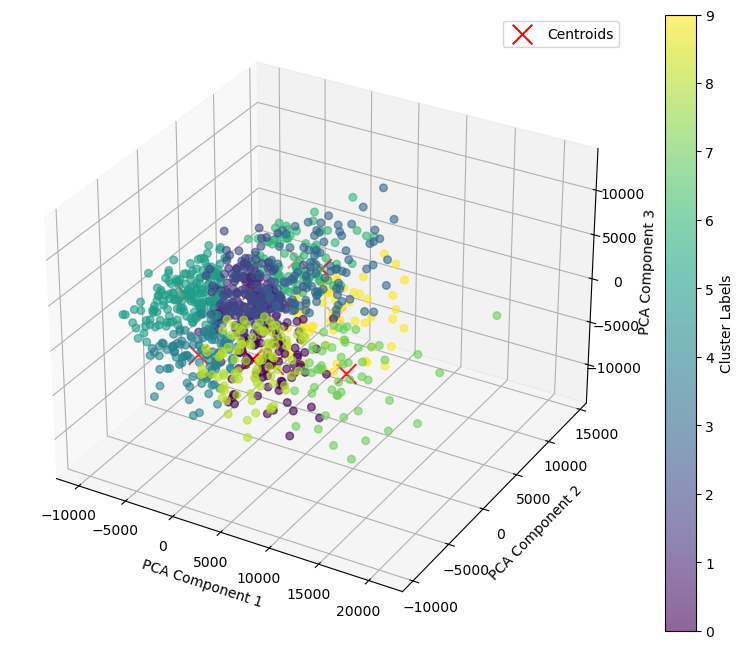}
    \caption{}
\end{subfigure}
\begin{subfigure}{0.49\linewidth}
    \centering
    \includegraphics[width=\linewidth]{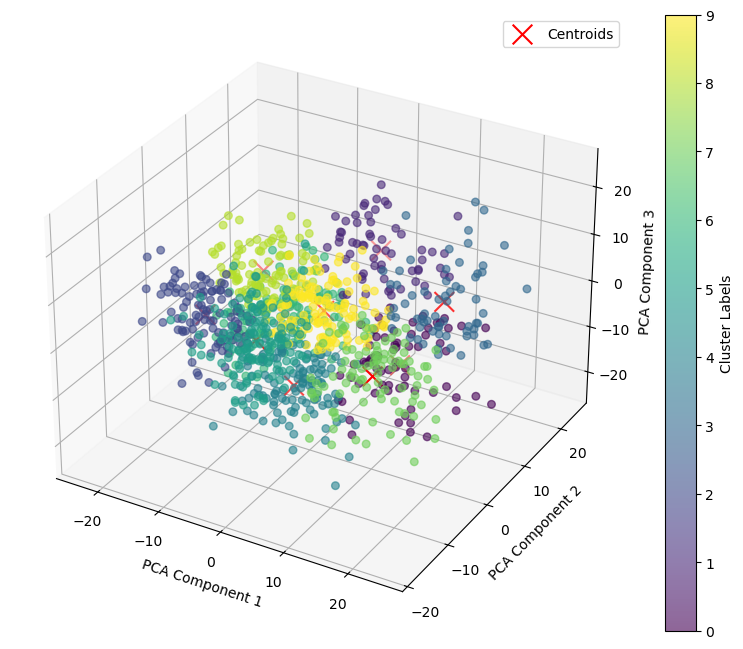}
    \caption{}
\end{subfigure}
\caption{K-means clustering results for class intermediate outputs from (a) VGG19 and (b) MobileNet.}
\label{fig:Class_plots}
\vspace{-0.2in}
\end{figure}

It is important to highlight that the classes are not clearly distinguishable based on the intermediate outputs collected from the layers of VGG and MobileNet, as shown in Figure \ref{fig:Class_plots}. The Silhouette Score for VGG is 0.24, and for MobileNet, it is 0.25, indicating a low level of separation between the categories. This lack of distinction suggests that the features extracted by these layers do not provide sufficient separation between the categories. One possible explanation is that these layers may capture general features instead of specific details related to each class, resulting in overlap within the feature space.

\subsection{Latent Space Manipulation using VAE}
\begin{figure}[t]
\centering
\begin{tabular}{ccccc}
\includegraphics[width=0.075\textwidth]{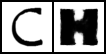} &
\includegraphics[width=0.075\textwidth]{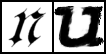} &
\includegraphics[width=0.075\textwidth]{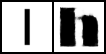} & \includegraphics[width=0.075\textwidth]{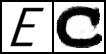} &
\includegraphics[width=0.075\textwidth]{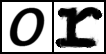} \\

\includegraphics[width=0.075\textwidth]{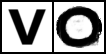} &
\includegraphics[width=0.075\textwidth]{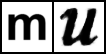} &
\includegraphics[width=0.075\textwidth]{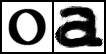} & \includegraphics[width=0.075\textwidth]{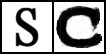} &
\includegraphics[width=0.075\textwidth]{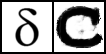} \\

\includegraphics[width=0.075\textwidth]{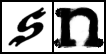} &
\includegraphics[width=0.075\textwidth]{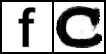} &
\includegraphics[width=0.075\textwidth]{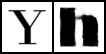} & \includegraphics[width=0.075\textwidth]{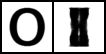} &
\includegraphics[width=0.075\textwidth]{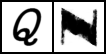} \\
\end{tabular}
\caption{Samples generated using one of the 10 farthest latent points, with each pair showing input (left) and transformed output (right).
}
\label{fig:fonts}
    \vspace{-0.2in}
\end{figure}

Next, we evaluate VAE's potential for misclassification attacks by training it on the Character Font Images dataset~\cite{character_font_images_417} and attempting to change the label by modifying the sample in the latent space. To achieve this, we divide the dataset into training and test sets with an 80-20 split. 
The VAE is trained with an encoder consisting of three layers: the first fully connected layer maps the flattened input (50$\times$50 image) to a 1000-dimensional space, followed by two additional layers that output the mean and log-variance of a 32-dimensional latent distribution. The decoder has two layers: the first fully connected layer maps the 32-dimensional latent vector to a 1000-dimensional space, and the second layer reconstructs the input back to a 2500-dimensional space. The VAE is trained with three encoder layers of sizes 256, 128, and 64, and a decoder with layers of sizes 64, 128, and 256, while preserving the distribution of samples in latent space. Next, we map the test samples into the latent space. Instead of using the latent space to create a similar sample, we calculate the distance between points in latent space using the Kullback–Leibler divergence between the learned distribution 
and a standard normal prior, and randomly select one of the 10 farthest points from the given test point to increase the likelihood of misclassification. Finally, we add noise and feed the modified latent vector into the decoder to generate a new sample. The qualitative results, shown in Figure~\ref{fig:fonts}, demonstrate that this method can visually transform one digit into a completely different one, leading to potential misclassification.

\section{Threat Landscape and Model}\label{sec:attackmodel}
In this section, we discuss the rationale and intent behind misclassification attacks on collaborative DNN inference, as well as the details of the system model and AdVAR-DNN attack methodology.

\subsection{Exploits of Misclassification} 
In general, IoT devices are vulnerable to cyberattacks because of their low-cost production, power constraints, connectivity issues, processing capabilities, data storage, and inherent heterogeneity since they involve a mix of network topologies,
hardware platforms, and servers. 
Data leakage and falsification are some of the most critical vulnerabilities that exist in different layers of a collaborative IoT-edge environment, viz., application, middleware, and edge layers \cite{mishra2021internet}, steaming for a varity of issues ranging from faulty codes, supply chain issues, malicious code injection, and other active attacks to name a few~\cite{
sadhu2022internet}. 
For instance, in the application layer, IoT devices rely on software often written in unsafe programming languages and poorly maintained due to their limited computational and power resources. Additionally, the hardware used in these devices is not always robust enough to withstand active attacks. This lack of robustness makes it easier for attackers to compromise a device within a network and utilize it as a base to launch attacks against other devices in the network \cite{meneghello2019iot}
The primary type of data falsification that can manifest through the IoT middleware layer, where weaknesses in data transmission and processing allow adversaries to inject false information 
or reconstruct input data and extract sensitive information. 
One of the most effective ways is through some form of Man-In-The-Middle (MITM) attack. Typical MITM attacks target data exchanged between endpoints, maybe remotely or often at intermediate hops/relay nodes through proxy devices, compromising both data integrity and confidentiality 
To this end, how such falsification can be performed while the intermediate output of a layer is transmitted from the IoT device to a remote server is beyond the scope of this paper. Instead, we focus on exploring if confidentiality and integrity violations can be performed on such data, and what manipulations can be carried out by the attackers to initiate misclassification.   

\begin{figure*}[t]
\centering
\includegraphics[width= 0.8 \textwidth]{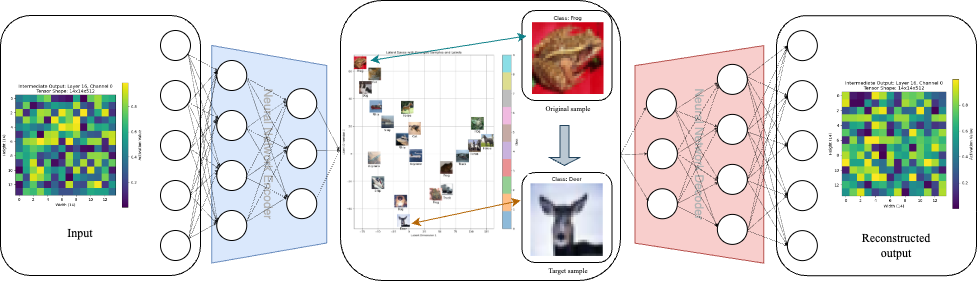}
\caption{
The overall workflow of the AdVAR-DNN attack. The adversary perturbs the extracted latent features before feeding them into the decoder, leading to the generation of adversarial samples that deviate significantly from the original input data.}
\label{fig:Workflow}
\vspace{-0.1in}
\end{figure*}


\subsection{Collaborative DNN Inference System Model}
We assume a collaborative DNN inference of object classification DNNs involving an IoT device and an edge/cloud server, possibly connected via one or many relay nodes/devices. In this setup, the same pre-trained target model is divided into two sequential components: \( M_1 \) and \( M_2 \). The input to the system, represented as \(x\), is an image that is initially processed by the IoT device using shallow layers of the model \( M_1 \). This stage extracts features from the input, producing an intermediate activation output, \( h \). The intermediate output \( h \) is then transmitted to the remote server, which processes it using the later layers of the model, \( M_2 \) to refine the features and generate the final classification result and a confidence metric for the object classification task. The object of the system is to classify objects with a high degree of confidence. We assume that a confidence value that is too low may trigger anomaly detection by the system.
The choice of the cut-point layer to divide the target model between the IoT device and the server can vary during the inference process. As explained before, the optimal cut point \(n\) determination is an active area of research which can be based upon optimizing a multitude of performance metrics, as explained in works, such as \cite{zhang2023effect, edgeRL, mounesan2025infer}.
We assume that both IoT devices and the remote servers are trusted and that an attacker gains access to the intermediate output data `in transit', i.e., after the data leaves the IoT device and before it reaches the server. Further, the attacker aims to cause misclassification with high degree of confidence (by the DNN) in order to avoid triggering system anomaly detection.

\subsection{AdVAR-DNN Attack Methodology}
\subsubsection{Attacker capabilities} 
For the collaborative DNN inference system model, we assume that the attacker can only access the intermediate activation output, \(h\), transmitted from the IoT device to the remote server. This assumption represents a minimal knowledge threat model, where the attacker has the least amount of information about the system. Specifically, the adversary has no access to the input image, \(x\), or the internal parameters of the pre-trained model and its partitions, including both \(M_1\)and \(M_2\). Additionally, the attacker does not know the identity of the cut-point layer, i.e., it is unaware of how many layers are in \(M_1\)and \(M_2\). 
Despite these constraints, we will demonstrate how AdVAR-DNN can manipulate \(h\) to deceive the original DNN model 
to misclassify unknown inputs.

\subsubsection{Workflow of Generating Adversarial Examples}
For AdVAR-DNN, we consider that the attacker can 
passively monitoring and collecting intermediate layer output over a long time. The resulting dataset, denoted as \( \mathcal{D}_h = \{ h_1, h_2, \dots, h_N \} \), consists of intermediate activation outputs for training the VAE that the attacker uses to manipulate latent representations and generate out-of-class samples through interpolation with the training latent space. 
Specifically, the attacker maps the collected dataset to the latent space using a function \( g: \mathcal{D}_h \to \mathcal{Z}\), where \(\mathcal{Z}\) represents the latent space. A latent vector \( z \sim P(Z)\)) is then sampled from the learned distribution and used by the VAE generative model to generate a new in-class object. While launching the attack, the attacker replaces the latent representation of the original sample, \(z_o\), which is meant to be sent to the server for the collaborative inference classification process, with the latent representation of the target, \(z_t \). The modified latent representation is then passed to the decoder of the VAE, denoted as \( D(\cdot) \), which generates a new sample:
\[
x' = D(z_t)
\]
where \(x'\) represents the generated sample, which is significantly different from the original input \( x_o = D(z_o) \).

Since the attacker selects a latent representation \(z_t\) from the learned space, the generated sample will likely be a meaningful representation of the data distribution rather than just random noise. The continuity of the latent space ensures that any small perturbation in $z$ results in meaningful outputs. Therefore, although the attacker has no access to the model parameters or the exact identity of the cut-point layer, the generated sample \( x' \) remains realistic and valid within the target class. Figure \ref{fig:Workflow} illustrates the overall workflow of the process, and shows how the attacker manipulates the intermediate representations to misguide the DNN model in generating faulty classification. 

\subsubsection{ Interpolating the Latent Space to Control Intermediate Representations} 

To control intermediate representations, the attack uses popular linear interpolation (Lerp) \cite{liu2018data} to create smooth transitions between two specific points within the latent representation of the original intermediate output and the target latent representation that will be decoded by the VAE to generate a new sample. The implementation of linear interpolation is straightforward and introduces minimal computation. The interpolation can be represented as follows: 
\[
z_{\alpha} = (1 - \alpha) z_o + \alpha z_t
\]
where $z_o$ and $z_t$ are the latent representations of the original and target data respectively, and $ \alpha \in [0,1] $ is the coefficient for linear interpolation such that when $ \alpha$ is close to 0, the sample is more similar to the original sample, and when $\alpha$ is close to 1, the sample closely resembles the target. Utilizing this method,  the attacker can control the intensity of the attack while remaining stealthy, as having varied perturbations in latent space may lead to an in-class sample.
The interpolation threshold ($\alpha$),  which also represents the attack strength, dictates how much the manipulated representation deviates from the original, which directly affects the likelihood of being detected during the execution of \(M_2\) layers at the final stages of DNN inference at the remote server.
Additionally, we explore an alternate Spherical linear interpolation (Slerp) \cite{white2016sampling} method 
for a more natural transition. The comparison between the two techniques will be discussed in Section~\ref{sec:evaluation}. 

\section{Evaluation and Results}
\label{sec:evaluation}

In this section, we evaluate the effectiveness of the proposed AVEA attack through an end-to-end experimental setup. As described in Section~\ref{sec:attackmodel}, the attack consists of two phases: (1) \textit{the collection phase}, where the attacker passively collects intermediate layer outputs, and (2) \textit{the active attack phase}, where the attacker uses a trained VAE to manipulate the collected features and generate adversarial samples. During the eavesdropping phase, we extract intermediate activations from 3 specific object classification models, viz., AlexNet, VGG19, and MobileNet, at the layers listed in Table~\ref{tab:cut_layers}. The data is collected from CIFAR-100 across 50,000 samples and used to construct a dataset for training a VAE for each model-layer pair. Although the number of samples for training is large, we later explore the optimal number of samples required to achieve the desired attack success. In the active attack phase, the trained VAE is leveraged to perturb the intermediate representations and generate adversarial samples, as described in Section~\ref{sec:attackmodel}.


For the end-to-end evaluation, the test data is first processed by the initial layers of a DNN running on the IoT device. The corresponding VAE modifies the intermediate representation at the designated cut-point layer to induce a misclassification before being received by the edge server. The manipulated representation is subsequently passed to the remaining layers of the model hosted on an edge server, and the final classification output is compared against the ground truth to assess attack effectiveness. The impact of the attack is quantified by the degradation in model accuracy.
Our objective is to analyze the model response and assess its vulnerability to adversarial samples generated through VAE. In the following sections, we present the experimental setup and results, demonstrating the efficacy of the proposed AdVAR-DNN attack through extensive evaluations.



\subsection{Experiment Setup}
We begin by outlining the used DNN models, the VAE architecture and the details of training.

\textbf{DNN Models:} We focus on three widely used DNNs for object classification tasks: AlexNet \cite{krizhevsky2017imagenet}, VGG \cite{sengupta2019going}, and MobileNet \cite{howard2017mobilenets}, all of which have been extensively studied in the context of collaborative DNN inference. For our experiments, we fine-tune the pre-trained versions of these models using transfer learning 
, adapting them to the CIFAR-100 dataset to achieve high accuracy. Table \ref{tab:cut_layers} provides an overview of the DNN models used in this study, along with their benchmarked accuracy, the corresponding cut point layers, and their output sizes. Although the accuracy of the models could be further optimized through additional fine-tuning, but it is not the primary focus of this study.

\begin{table}[t]
    \centering
    \caption{DNN Models, Their baseline accuracy, Selected Cut Point Layers, and Corresponding Output Sizes of Cut Point Layers}
    \label{tab:cut_layers}
    \begin{tabular}{p{1.1cm}p{1.5cm}p{2cm}p{2cm}}
        \hline
        \textbf{Model} & \textbf{Accuracy}&  \textbf{Cut Layer} & \textbf{Output Size} \\  
        \hline
        \multirow{4}{*}{AlexNet} &\multirow{4}{*}{77.97\%}& 3 & \(27 \times 27 \times 192\) \\  
         && 6 & \(13 \times 13 \times 384\) \\  
         && 8 & \(13 \times 13 \times 256\) \\  
         && 10 & \(13 \times 13 \times 256\) \\  
        \hline
        \multirow{4}{*}{VGG19} &\multirow{4}{*}{82.2\%}& 12 & \(28 \times 28 \times 512\)\\  
         && 16 & \(14 \times 14 \times 512\) \\  
         && 18 & \(14 \times 14 \times 512\)\\  
         && 20 & \(14 \times 14 \times 512\) \\  
        \hline

        \multirow{4}{*}{MobileNet} &\multirow{4}{*}{69.4\%}& 20 & \(56 \times 56 \times 128\) \\  
         && 40 & \(14 \times 14 \times 512\) \\  
         && 50 & \(14 \times 14 \times 512\) \\  
         && 63 & \(14 \times 14 \times 512\) \\  
        \hline
    \end{tabular}
    \vspace{-0.2in}
\end{table}

\textbf{VAE Architecture:} As mentioned earlier, the distributions of the intermediate features for each model and cut-point layer differ. As a result, for each layer of the specific model, the attacker trains a specific VAE. In this section, we describe the architecture of an exemplary VAE designed and trained for input values with dimensions $(14,14,512)$.
The VAE encoder consists of two fully connected hidden layers that process the input and map it to the latent space. The first hidden layer reduces the input dimensionality from (14, 14, 512) to a hidden size of 1000 neurons, but it may change for other cases as part of fine-tuning. Similarly, the latent space size is also fine-tuned, and for this specific example, it is set to 32. This is followed by two separate layers that learn the mean and log variance of the latent distribution, with a latent space size of 32.  The input size changes based on the cut layer, as it is shown in Table \ref{tab:cut_layers}. The decoder mirrors this structure, where the first layer transforms the latent vector of size 32 into a higher-dimensional space, followed by another fully connected layer that reshapes it back to the input feature size (14, 14, 512) corresponding to the cut layer. A sigmoid function is applied at the end to ensure the output values remain within a valid range, facilitating the generation of meaningful adversarial samples.
Our VAE model is implemented in PyTorch. We first train the VAE using the data collected by the attacker, employing the ADAM optimizer \cite{kingma2014adam}, and save the trained model for subsequent use in generating adversarial samples.
The model is trained with a loss function that combines \textit{Mean Squared Error (MSE)} reconstruction loss and KL Divergence. The reconstruction loss measures the difference between the input and the reconstructed output, while the KL divergence encourages the mean and variance pairs of the latent variables to remain close to a standard normal distribution \cite{luo2018adversarial}.

\textbf{Dataset:} We utilize the CIFAR-100 dataset to fine-tune the subject DNN models and evaluate the effectiveness of our attack by generating adversarial examples. CIFAR-100 contains 100 classes, each with 600 images. The dataset is split such that 83.33\% of the images are used for training and 16.67\% for testing, resulting in 50,000 training images and 10,000 test images. These 100 fine-grained classes are grouped into 20 broader superclasses. All images have a resolution of 32$\times$32 pixels and three color channels (RGB) \cite{krizhevsky2009learning}.

\begin{table}[t]
\centering
\caption{Visualization of selected samples comparing original images vs. generated adversarial examples using intermediate features of layer 20 in the VGG19 model for different values of $\alpha$.}
\resizebox{\columnwidth}{!}{%
    \begin{tabular}{m{2.5cm}c c c c c c }
        \hline
        & \multicolumn{6}{c}{\large Assigned class and confidence} \\
        \hline
        \noalign{\vskip 1pt}
        \textbf{\normalsize Class / Conf.} &
        \normalsize Human (0.73) & \normalsize Bridge (0.99) & \normalsize Orchid (0.43) & \normalsize Truck (0.97) & \normalsize Elephant (0.74) & \normalsize Lizard (0.96) \\
        \cdashline{0-0} 
        \raisebox{40pt}{\textbf{ \small Original Model}} &
        \includegraphics[width=0.1\textwidth]{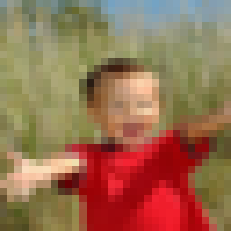} &
        \includegraphics[width=0.1\textwidth]{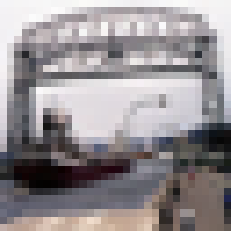} &
        \includegraphics[width=0.1\textwidth]{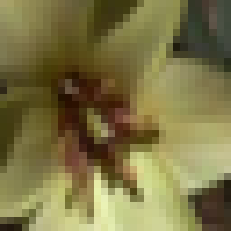} &
        \includegraphics[width=0.1\textwidth]{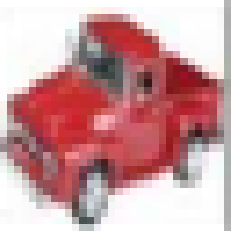} &
        \includegraphics[width=0.1\textwidth]{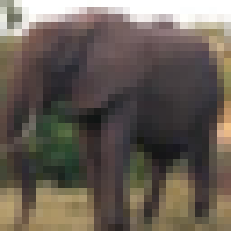} &
        \includegraphics[width=0.1\textwidth]{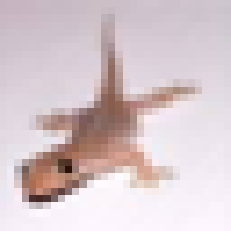} \\
        
        \hline 
        \noalign{\vskip 1pt}
        \textbf{\normalsize Class / Conf.} &
        \normalsize Human (0.59) & \normalsize Bridge (0.99) & \cellcolor{pink}  {\normalsize Rose (0.78) }& \normalsize Truck (0.52) & \normalsize Elephant (0.99) &  \cellcolor{pink}{\normalsize Lizard (0.49)} \\
        \cdashline{0-0} 
        \raisebox{40pt}{\textbf{\small Non-malicious VAE}} & 
        \includegraphics[width=0.1\textwidth]{Figures/sample_visualization/1.png} &
        \includegraphics[width=0.1\textwidth]{Figures/sample_visualization/2.png} &
        \includegraphics[width=0.1\textwidth]{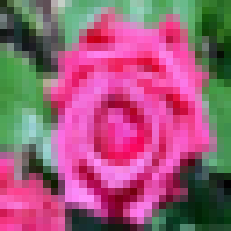} &
        \includegraphics[width=0.1\textwidth]{Figures/sample_visualization/4.png} &
        \includegraphics[width=0.1\textwidth]{Figures/sample_visualization/5.png} &
        \includegraphics[width=0.1\textwidth]{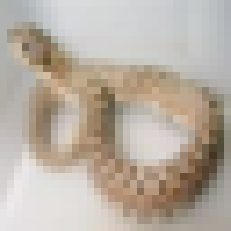} \\
        
        \hline
        \noalign{\vskip 1pt}
        \textbf{\normalsize Class / Conf.} &
        \cellcolor{pink} {\normalsize Keyboard (0.34)} & \cellcolor{pink}{ \normalsize Rocket (0.79)} & \cellcolor{pink}{ \normalsize Skyscraper (0.96)} & \cellcolor{pink}{\normalsize Trout (0.45)} & \cellcolor{pink}{ \normalsize Orange (0.71)} & \cellcolor{pink}{\normalsize Couch (0.42)} \\
        \cdashline{0-0} 
        \raisebox{40pt}{\textbf{\small AdVAR-DNN samples}} & 
        \includegraphics[width=0.1\textwidth]{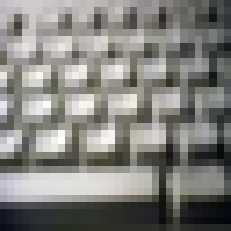} &
        \includegraphics[width=0.1\textwidth]{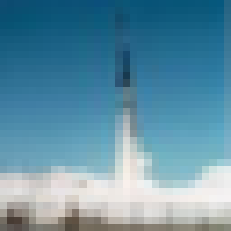} &
        \includegraphics[width=0.1\textwidth]{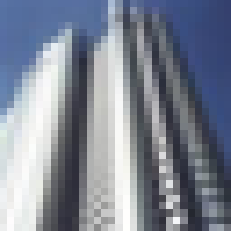} &
        \includegraphics[width=0.1\textwidth]{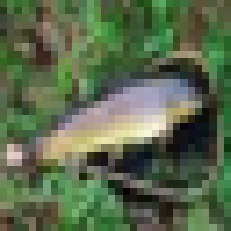} &
        \includegraphics[width=0.1\textwidth]{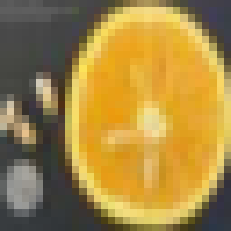} &
        \includegraphics[width=0.1\textwidth]{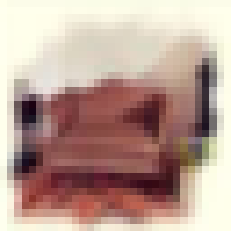} \\
    \end{tabular}
} 
\label{tab:visualization_label}
\end{table}


\textbf{Evaluation metrics}
We evaluate the performance of the attack using confidence, accuracy drop, and Attack Success Rate (ASR) as key metrics to measure attack effectiveness. 
\begin{itemize}[leftmargin=*,itemsep=0pt]
    \item \textbf{Accuracy} assesses the DNN model’s ability to classify inputs correctly. A successful attack sharply reduces accuracy, causing widespread misclassification of adversarial samples. 
    \item \textbf{Confidence} quantifies the probability assigned to the predicted class by the DNN model, indicating its confidence in the prediction. A high score reflects strong certainty, while a low score suggests uncertainty. This is crucial in misclassification attacks, where adversaries aim to induce high-confidence errors that remain untraceable.
    \item \textbf{Attack Success Rate (ASR)} 
    evaluates how often the attacker successfully degrades the DNN models' classification performance. It shows the percentage by which accuracy decreases in comparison to the baseline accuracy, which is the DNN model's accuracy on VAE-based generated samples before applying the proposed AdVAR-DNN attack ($\alpha = 0$). A higher ASR indicates that adversarial perturbations are significantly reducing classification performance, while a lower ASR suggests that the DNN models maintain some resilience even after the attack. 
\end{itemize}

\subsection{Attack Efficacy}
In the first set of experiments, we conduct a qualitative evaluation of selected samples from an end-to-end experiment to assess the effectiveness of the proposed AdVAR-DNN adversarial attack. Our objective is to analyze the DNN's susceptibility to these VAE-generated adversarial samples.
To achieve this, we feed samples from the test portion of CIFAR-100 dataset into the model and compare the assigned class and confidence scores under three conditions: 1) the original model without any attack, assuming that intermediate features reach the server without manipulation or perturbation; 2) non-malicious generative samples, where the intermediate representations are processed by a trained VAE to generate highly similar samples without inducing misclassification; and 3) adversarial samples, where values are manipulated in the latent space to trigger misclassification.

Table \ref{tab:visualization_label} presents qualitative results of VGG19 with cut-points at layer 20. 
We can observe that non-malicious VAE-generated samples that are mostly distortion-free, when processed by the \(M_2\) layers of the DNN, yield slightly lower accuracy than the baseline, indicating that the VAE generated high-quality samples resembling the originals.
This is due to the inherent limitations of reconstruction-based learning, which leads the VAE in struggling to create robust latent representations.
However, under the AdVAR-DNN attack, the DNN model accuracy drops to 0\%, signifying complete misclassification of the adversarial samples. Despite this, the model maintains high confidence in its incorrect predictions, highlighting the effectiveness of the attack and the VAE's ability to generate high-quality adversarial samples.

\subsection{Attack Scale} 
Next, we present quantitative DNN accuracy results 
under AdVAR-DNN attack.
Table \ref{tab:alex_attack_accuracy} summarizes the accuracy results of AlexNet, where the model achieves a baseline accuracy of 77.97\%  after transfer learning on CIFAR-100. When the model is tested on VAE-generated samples with no attack (i.e., $\alpha = 0$), the accuracy drops across all layers. 
Under maximum attack strength (i.e., $\alpha = 1$), the model performance degrades further, with deeper layers (Layers 8 and 10) reaching an accuracy of 0.0\% due to strong perturbations. 
Table \ref{tab:VGG_attack_accuracy} demonstrates a similar accuracy trajectory, this time for  VGG19, where accuracy drops sharply with increased $\alpha$.
As the attack strength reaches $\alpha = 1$, the model accuracy decreases furthest, especially in deeper layers.
Finally, in Table \ref{tab:Mobilenet_attack_accuracy}, we observe that, unlike AlexNet and VGG19, MobileNet's accuracy takes a huge dip even without adversarial perturbation (i.e., $\alpha = 0$). This suggests that MobileNet is highly sensitive to distribution shifts caused by VAE-based generated samples. This sensitivity may indicate that the VAE fails to generate meaningful samples that align with MobileNet’s learned feature representations. This further suggests that different DNNs have different levels of sensitivity to perturbations, as models, such as MobileNet, struggle to classify VAE-based generated data even before any adversarial perturbations are applied, which makes adversarial intent more detectable for such DNNs.

\begin{table}[h]
    \centering
    \caption{Impact of VAE and Adversarial Attacks on Model Accuracy for AlexNet}
    \label{tab:alex_attack_accuracy}
    \resizebox{\columnwidth}{!}{
    \begin{tabular}{cccc}
        \toprule
        \textbf{Layer} & \textbf{Baseline Accuracy} & \textbf{Accuracy at $\alpha = 0$}  & \textbf{Accuracy at $\alpha = 1$}  \\
        \midrule
        Layer 3  & 77.97\%  & 13.0\%   & 3.0\%  \\
        Layer 6  & 77.97\%  & 44.0\%   & 1.0\%  \\
        Layer 8  & 77.97\%  & 36.0\%   & 0.0\% \\
        Layer 10 & 77.97\%  & 41.0\%   & 0.0\%  \\
        \bottomrule
    \end{tabular}
    }
\end{table}

\begin{table}[h]
    \centering
    \caption{Impact of VAE and Adversarial Attacks on Model Accuracy for VGG19}
    \label{tab:VGG_attack_accuracy}
    \resizebox{\columnwidth}{!}{
    \begin{tabular}{cccc}
        \toprule
        \textbf{Layer} & \textbf{Baseline Accuracy} & \textbf{Accuracy at $\alpha = 0$}  & \textbf{Accuracy at $\alpha = 1$} \\
        \midrule
         Layer 12  & 82.2\%  & 19.0\%  & 1.0\%   \\
        Layer 16  & 82.2\%  & 50.06\%   & 42.45\%   \\
        Layer 18  & 82.2\%  & 67.0\%   & 0.0\%  \\
        Layer 20  & 82.2\%  & 78.0\%  & 0.0\%  \\
        \bottomrule
    \end{tabular}
    }
\end{table}

\begin{table}[h]
    \centering
    \caption{Impact of VAE and Adversarial Attacks on Model Accuracy for MobileNet}
    \label{tab:Mobilenet_attack_accuracy}
    \resizebox{\columnwidth}{!}{ 
    \begin{tabular}{cccc}
        \toprule
        \textbf{Layer} & \textbf{Baseline Accuracy} & \textbf{Accuracy at $\alpha = 0$} &  \textbf{Accuracy at $\alpha = 1$} \\
        \midrule
        Layer 20  & 69.4\%  & 2.0\%  & 1.0\%   \\
        Layer 40  & 69.4\%  & 4.0\%  & 0.0\%   \\
        Layer 50  & 69.4\%  & 15.5\%   & 0.0\%   \\
        Layer 63  & 69.4\%  & 7.84\%  & 1.0\%  \\
        \bottomrule
    \end{tabular}
    } 
\end{table}

\begin{figure}[t]
    \centering
    \begin{subfigure}{0.22\textwidth}
        \centering
        \includegraphics[width=\textwidth]{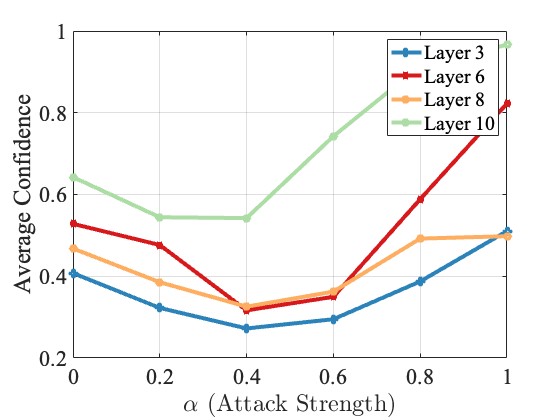}
        \caption{}
        \label{fig:Alex-confidence}
    \end{subfigure}
    \begin{subfigure}{0.22\textwidth}
        \centering
        \includegraphics[width=\textwidth]{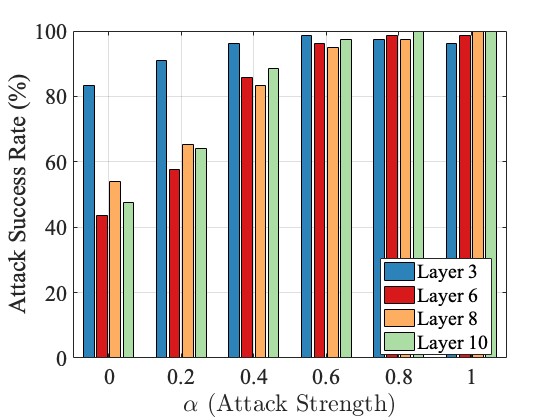}
        \caption{}
        \label{fig:Alex-ASR}
    \end{subfigure}
    \caption{Performance of AlexNet under adversarial attack vs.$\alpha$ (attack strength) : (a) DNN model accuracy degradation, (b) DNN model confidence of prediction, (c) Attack Success Rate (ASR)}
    \label{fig:Alex_combine}

\end{figure}

Figure \ref{fig:Alex_combine} illustrates AlexNet performance in terms of average classification confidence and AdVAR-DNN attack success rate (ASR) for varying $\alpha$.
Figure \ref{fig:Alex-confidence} presents the impact of the attack on classification confidence when the attack uses output from different intermediate layers. At $\alpha = 0$, the AlexNet classification confidence is low when AdVAR-DNN uses the output of layer 3. This is because layer 3 outputs low-level features that make it difficult for VAE to learn an effective compressed latent space. As a result, the VAE-generated samples are less meaningful to AlexNet. In contrast, intermediate data collected from the deeper layers capture better features that enable the VAE to generate more structured samples, leading to higher confidence towards erroneous classification, especially for higher $\alpha$. 
However, a too large $\alpha$ can become suspicious and trigger the anomaly detection mechanism. 
An optimal attack strength can thus be selected in such a way that it provides a balance between stealthiness and high attack success.

Figure \ref{fig:Alex-ASR} shows ASR as a function of attack strength. We observe that at $\alpha = 0$, ASR is relatively low, which demonstrates that AlexNet classification performance degradation is noticeable even on VAE-generated samples. As $\alpha$ increases, ASR rises sharply, which indicates that the adversarial perturbation is able to degrade AlexNet model performance effectively. Earlier cut layers show a slower ASR increase in comparison to the deeper cut layers, which demonstrates that the early feature extraction layers maintain some robustness against VAE-based generative attacks.

\begin{figure}[t]
    \centering
    \begin{subfigure}{0.22\textwidth}
        \centering
        \includegraphics[width=\textwidth]{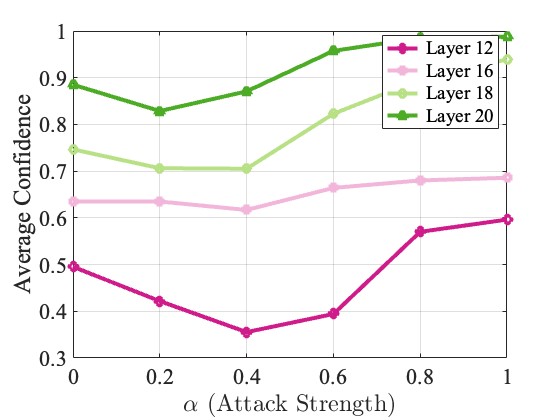}
        \caption{}
        \label{fig:VGG-confidence}
    \end{subfigure}
    \begin{subfigure}{0.22\textwidth}
        \centering
        \includegraphics[width=\textwidth]{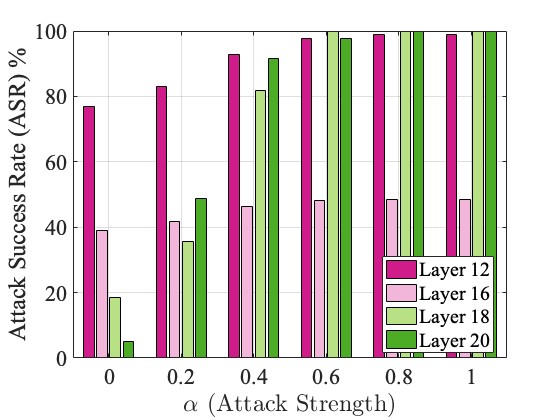}
        \caption{}
        \label{fig:VGG-ASR}
    \end{subfigure}
    \caption{Performance of VGG under adversarial attack vs.$\alpha$ (attack strength) : (a) DNN model accuracy degradation, (b) DNN model confidence of prediction, (c) Attack Success Rate (ASR)}
    \label{fig:VGG_combine}
    \vspace{-0.1in}
\end{figure}

Figure \ref{fig:VGG_combine} illustrates VGG19 performance for varying $\alpha$.
Similar to AlexNet, VGG19 shows that with an increasing $\alpha$, classifier confidence in correct classification drops significantly, and while confidence in incorrect classification increases after a certain point, as seen in Figure \ref{fig:VGG-confidence}. This overconfidence in misclassification gives the attacker assurance that the VGG19 model is not only misclassifying but also remains highly confident in its decisions, making attack detection difficult.
VGG19 ASR results shown in Figure \ref{fig:VGG-ASR} demonstrate behavior similar to that of AlexNet, where ASR reaches almost 100\% at $\alpha=0.6$ for layers 18 and 20. This demonstrates that at higher attack strengths, all adversarial samples can effectively deceive VGG19, resulting in complete misclassification.

\begin{figure}[t]
    \centering
    \begin{subfigure}{0.22\textwidth}
        \centering
        \includegraphics[width=\textwidth]{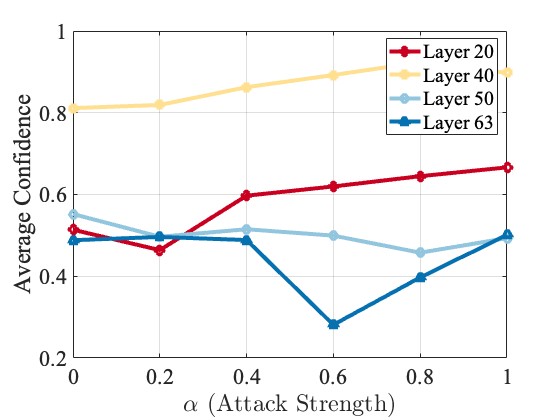}
        \caption{}
        \label{fig:Mnet-confidence}
    \end{subfigure}
    \begin{subfigure}{0.22\textwidth}
        \centering
        \includegraphics[width=\textwidth]{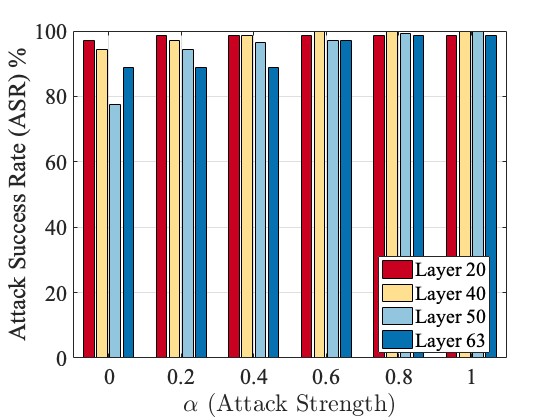}
        \caption{}
        \label{fig:Mnet-ASR}
    \end{subfigure}
    \caption{Performance of MobileNet under adversarial attack vs.$\alpha$ (attack strength) : (a) Classifier accuracy degradation, (b) Classifier confidence of prediction, (c) Attack Success Rate (ASR)}
    \label{fig:Mnet_combine}
    \vspace{-0.1in}
\end{figure}

Figure \ref{fig:Mnet_combine} shows that, unlike AlexNet and VGG19, MobileNet exhibits a different response to AdVAR-DNN attack. 
Similar to its accuracy behavior in Table~\ref{tab:Mobilenet_attack_accuracy}, 
MobileNet classifier confidence across different $\alpha$ remains very low, demonstrating model uncertainty towards VAE-generated samples even before adversarial perturbations are introduced.
The low initial confidence shows that the generated samples from intermediate features of MobileNet are more detectable and may trigger anomaly detection. However, if the attacker's objective is more brute force than stealthy, the result from Figure \ref{fig:Mnet-ASR} proves that such an approach can be effective, as for any $\alpha > 0.6$, ASR exceeds 90\% for all layers.

\begin{figure}[h]
    \centering
    \begin{subfigure}{0.23\textwidth}
        \centering
        \includegraphics[width=\linewidth]{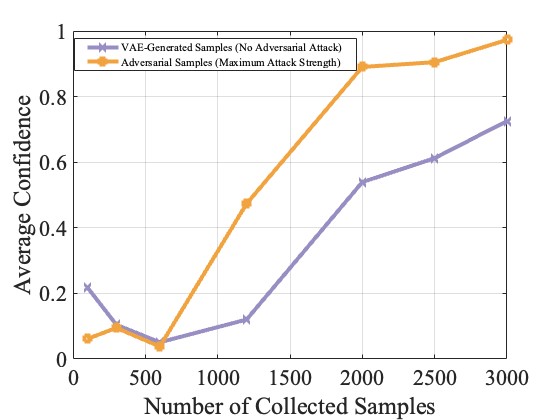}
        \caption{}
        \label{fig:confidence_samples}
    \end{subfigure}
    \hspace{0.1cm}
    \begin{subfigure}{0.23\textwidth}
        \centering
        \includegraphics[width=\linewidth]{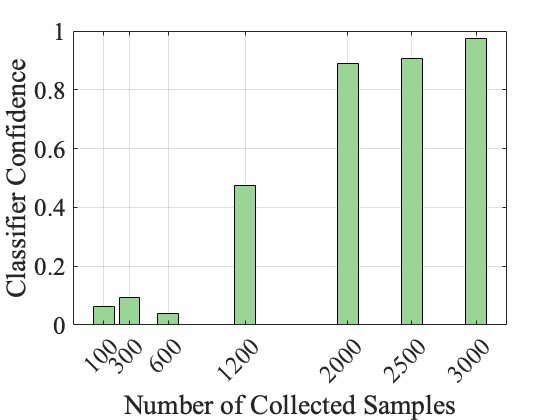}
        \caption{}
        \label{fig:confidenceBar_alpha1}
    \end{subfigure}
    \vspace{-0.1in}
    \caption{Impact of Intermediate Collected Samples on VAE Training and DNN model Deception (VGG19 Layer 20). Increasing the number of collected samples leads to higher DNN model confidence in misclassified predictions.}
    \label{fig:vgg19_growing samples}
    \vspace{-0.2in}
\end{figure}

\begin{figure}[t]
    \centering
    \begin{subfigure}{0.23\textwidth}
        \centering
        \includegraphics[width=\textwidth]{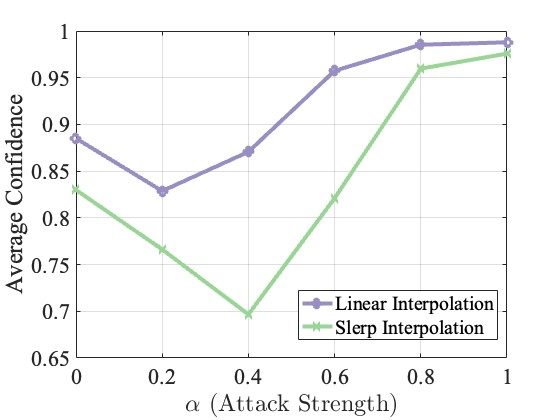}
        \caption{}
        \label{fig:Interpolation-confidence}
    \end{subfigure}
    \begin{subfigure}{0.23\textwidth}
        \centering
        \includegraphics[width=\textwidth]{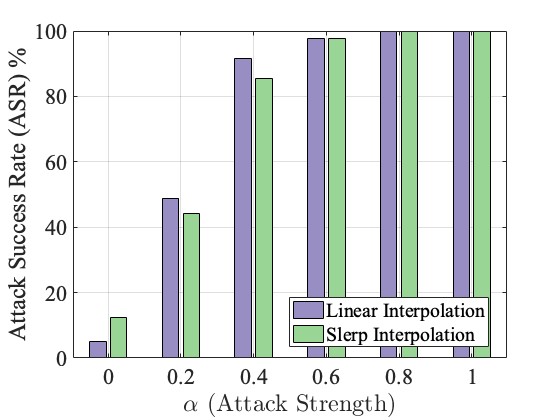}
        \caption{}
        \label{fig:Interpolation-ASR}
    \end{subfigure}
    \vspace{-0.1in}
    \caption{
    Performance of VGG19 under adversarial attack using different interpolation techniques vs.$\alpha$ (attack strength) : 
    (a) Classifier confidence of prediction, (b) Attack Success Rate (ASR)
    }
    \label{}
    \vspace{-0.2in}
\end{figure}

\subsection{Attack Data Budget}
\label{sec:attack-data-budget}
The success of an adversarial attack depends on both the model architecture and the amount of data collected for training the VAE. To analyze this, we experiment with different degrees of data collection from the intermediate output of the cut-point layer. 



Figure \ref{fig:vgg19_growing samples} illustrates the effect of increasing the size of collected samples from the cut-point layer in terms of 
confidence of classification, and ASR.
Figure \ref{fig:confidence_samples} compares the DNN model (VGG19 in this case) confidence for VAE-generated samples with and without adversarial perturbations. 
The results reveal that the confidence in VAE-generated samples remains moderate and grows as more samples are collected.
We observe that for adversarial samples, the DNN model's confidence increases sharply at higher sample sizes. This trend continues until a sample size of 2000, beyond which the improvement in confidence somewhat settles. 
Finally, Figure \ref{fig:confidenceBar_alpha1} illustrates the DNN model confidence when it is exposed to adversarial data generated from different amounts of collected samples at maximum attack strength ( $\alpha = 1$). The result corroborates with Figure \ref{fig:confidence_samples}, where the confidence increases beyond sample size 2000 is marginal. This analysis provides key insights into the optimal sample size for an effective attack without making it dependent on oversampling with diminishing return, thus making the approach more realistic.


\subsection{Effect of Interpolation on AdVAR-DNN Attack}
Next, we compare Spherical Linear Interpolation (Slerp) and Linear Interpolation towards attack effectiveness. The comparison is conducted in terms of 
classification confidence and ASR on the VGG19 at the cut-point layer 20.



From the classification confidence performance in Figure~\ref{fig:Interpolation-confidence}, 
we observe that linear interpolation maintains a higher confidence level for all attack strengths and reaches to 0.9879 at $\alpha = 1$ while Slerp shows lower confidence at moderate attack strength. Finally, Figure \ref{fig:Interpolation-ASR} presents the ASR trends 
where it rises as the attack strength increases, with both methods reaching near-total misclassification at $\alpha  = 0.8$. We observe that the linear Interpolation achieves higher ASR at lower $\alpha$ values, indicating a stronger attack impact in early stages. On the other hand, Slerp maintains a more controlled increase in ASR but eventually converges to similar levels as linear interpolation. In conclusion, choosing one of these methods can be based on the attack's objective, whether it is to minimize model degradation quickly or maintain control over attack effectiveness and stay stealthy. All evaluation related codes and data are available through Github~\cite{git-lcn}.

\section{Discussions and Conclusions}
\label{sec:conclusion}
In this paper, we proposed AdVAR-DNN, a VAE-based adversarial attack on collaborative DNN inference involving IoT devices to artificially induce misclassifications. First, using benchmarking experiments, we showed how it is non-trivial for an attacker to generate adversarial samples for intermediate cut-point outputs and then how a VAE can help in manipulating such in a way that can trigger misclassification with high certainty, thereby avoiding suspicion. With detailed experimental evaluation using the most popular classification DNNs, we showed how the proposed AdVAR-DNN attack achieved a high success rate with little probability of detection.

In this work, we found that one of the key factors that affected the performance of the AdVAR-DNN attack was the cut-point layer in the DNN models. To analyze this, we experimented with different cut-point layers that demonstrated how different layers impacted AdVAR-DNN performance and system resilience against such attacks. 
We found that deeper layers in neural networks compressed similar features within a class and improved class discrimination. They focused more on high-level concepts, such as complex patterns, rather than raw details like edges. However, this abstraction might make them more vulnerable to adversarial manipulation. Empirical studies have shown that feature compression increases for deeper layers, while shallow layers retain more low-level details, offering greater robustness to adversarial attacks \cite{wang2023understanding}.


Further, previous studies \cite{he2019model} have demonstrated that shallow layers in a collaborative DNN inference framework were more vulnerable to reconstruction attacks. In such a scenario, an adversary with access to the intermediate output of a DNN model, could reconstruct the input data using different scenarios including white-box setting where the adversary could use the model parameters to recover the input data, black-box setting where the adversary could gain knowledge about the model by querying the inference system. In contrast, our proposed attack focused on the effectiveness of VAE-based adversarial samples. Our findings showed that in the context of adversarial samples, deeper layers were more vulnerable in comparison to earlier layers. This difference underlined a fundamental difference in attackers' objectives and suggested that reconstruction attacks exploited visibility-critical features in shallower layers, and our proposed attack was able to manipulate the decision-critical features and made the deeper layers the primary target for misclassification attacks. For future, investigating effective countermeasures remains an important direction of further exploration.



\bibliography{reference}

\end{document}